\documentstyle[preprint,aps,psfig]{revtex}

\def\lsim{\raise0.3ex\hbox{$\;<$\kern-0.75em\raise-1.1ex
\hbox{$\sim\;$}}}
\def\gsim{\raise0.3ex\hbox{$\;>$\kern-0.75em\raise-1.1ex
\hbox{$\sim\;$}}}
\begin{document}

\begin{flushright}
TMUP-HEL-0007
\end{flushright}
\begin{center}
\Large\bf
Answering the Sphinx's Questions on Neutrinos
\end{center}
\vskip 0.5cm
\begin{center}
Hisakazu Minakata\footnote
{Talk presented at Workshop on Neutrino Oscillations and Their Origin, 
Fujiyoshida, Japan, February 11-13, 2000, to appear in Proceedings 
published by Universal Academy Press, Tokyo.}\\
{\it Department of Physics, Tokyo Metropolitan University \\
1-1 Minami-Osawa, Hachioji, Tokyo 192-0397, Japan, and \\
Research Center for Cosmic Neutrinos,
Institute for Cosmic Ray Research, \\ 
University of Tokyo, Kashiwa, Chiba 277-8582, Japan\footnote
{Address correction requested}
}
\end{center}

\begin{abstract}
In answering the difficult questions on neutrinos asked by Sphinx 
I argue that search for proton decay is the most important 
experiment in coming 5-10 years. 
I also emphasize the crucial importance of the neutrinoless 
double beta decay with sensitivity of 
$\langle m_{\nu e}\rangle \sim 0.01$ eV 
level as the unique feasible way of directly detecting neutrinos 
of atmospheric mass scale in laboratories. 
I point out that, if observed at this level, it means not only that 
neutrinos are Majorana particle but also that they must obey an 
inverted mass hierarchy. 
\end{abstract}
\newpage

As one of the panellers in this session, I was asked by Yoichiro 
Suzuki, the head of the Priority Area Project on Neutrino Oscillations 
and their Origins, to answer the questions 

\noindent
$Q_1$: "What would be the most important neutrino oscillation experiment 
in coming five years?"
and the other related ones. Clearly, he was so kind to raise such  
a difficult question to answer. Of course, his intention was quite 
right, aiming at triggering interesting discussions in the session. 
Therefore, let me try to answer it as far as I can do. 

Actually I will try to answer it by making it harder, I mean 
more general, not by making it more specific. 
(A famous mathematitian Kiyoshi Oka once claimed that one must let 
problem harder by making it more general to beautifully solve it.)
So the question I try to answer is:

\noindent
$Q_2$: "What would be the most important experiment in coming five years 
which has anything to do with neutrino masses? 

Since the questions are so heavy I had to ask myself about 
what we know with confidence and what we do not know for sure in 
particle physics. After some wondering I found that the answer is 
simple. The standard model of particle physics \cite {GWS}, and 
that is about it.
 
We do know that the standard model is a great 
theory which explains almost everything that we know; they mostly 
come from the experiments done at energies of less than 1 TeV. 
A bit of more thought, however, reveals that it has profound 
consequences on physics far beyond the electroweak scale.
It stems from the nature of the standard model of being a 
renormalizable theory. It predicts that baryon and lepton 
number nonconserving processes are largely suppressed. 
It is because the gauge invariance and the renormalizablity 
do not allow us to have lepton and baryon number violating 
interactions in the standard model \cite {W79}. 
These interactions are possible only as unrenormalizable, 
higher-dimensional operators, and hence are suppressed by powers 
of a large mass scale which signals the energy scale that 
characterizes opening of new physics.
One of the simplest possible terms that violates lepton and 
baryon numbers is
\begin{equation}
\frac{1}{M^2}
\bar{e}\bar{d}uu
\end{equation}
which would mediate proton decay, 
$p \rightarrow \pi^0 + e^+$.
This is the basic reason why the proton decay is so rare; 
The basic defining principle of the theory, 
the gauge invariance and the renormalizablity, themselves 
guarantee the stability of matter without recourse to any extra 
symmetries that can be fragile. The most stringent bound to date 
comes from Superkamiokande \cite{Shiozawa}.

It should be stressed that the same reasoning in the standard 
model implies that the neutrino masses must be tiny. The lowest 
possible dimension 5 operator for the neutrino mass term is;
\begin{equation}
\frac{1}{M}
\phi\phi\nu\nu
\label{numass}
\end{equation}
where $\phi$ indicates the Higgs field. We know that 
$\langle \phi \rangle$ is 
about 250 GeV. If we take $M \simeq 10^{16}$ GeV as is natural for 
grand unification \cite{GG74}, then it gives us for a neutrino mass, 
$m_{\nu} \simeq 6 \times 10^{-3}$ eV. 
Our prejudice in the hierarchy of lepton and quark masses 
then suggests that the largest $\Delta m^2$ is given by 
$m_{\nu}^2 \simeq 4 \times 10^{-5}$ eV, 
which is not so far from the $\Delta m^2$ scales implied by 
the atmospheric neutrino data and the MSW solar neutrino solutions. 
(Most probably it is in between them.)

Therefore, it appears to me that the most natural interpretation 
of the tiny neutrino mass which we observe by various neutrino 
experiments is the one {\it expected} by the most well-tested theory, 
the standard model. Its size naturally suggests the grand unification 
mass scale as the energy threshold for new physics. 
If this interpretation is correct, the next step is obvious; 
observation of proton decay. 

Thus, guided by my conservatism in which I trust only the generic 
features of the standard model as a renormalizable theory based 
on gauge principle, I was led to an answer to the modified version 
of Yoichiro's question $Q_2$; 
The most important experiment in coming 5-10 years is the search 
for proton decay.
I have no reason not to suspect that it will be observed by 
Superkamiokande; I do not share the pessimism which apparently 
possessed by the experimentalists. I hope that they are patient 
enough to continue to believe in this generic "prediction" of 
the standard model, albeit not in various GUT-model-dependent 
predictions on dominant modes.

Now, since this is the Neutrino Workshop I feel obliged to address 
at least one neutrino experiment which I believe to be of key 
importance in the near future. 
It is the neutrinoless double beta decay experiment \cite {DKT}. 
There are at least two impressive bold attempts to reach 
sensitivity of $\sim$ 0.01 eV \cite{GENIUS,Moon}. Because it is 
smaller than square root of the atmospheric $\Delta m^2$, 
we have a good chance of observing real events, 
not just placing the bound, if the neutrinos are Majorana particles 
and if the experiments are feasible. 
It should be emphasized that it is the unique experiment, 
to my knowledge, that is capable of proving of Majorana 
nature of neutrinos.  
I note that the extra Majorana phases cannot be observable in 
any neutrino oscillation experiments in vacuum and in matter. 

I would be very happy if I can stop here. But I must point out 
the following fact, though it might give a little harder time 
for experimentalists. 
What I told you a moment ago was not quite correct. To clarify 
what I mean by this we have to distinguish the normal 
($m_3 \gg m_1 \sim m_2$) and the inverted ($m_1 \sim m_2 \gg m_3$)
hierarchies of neutrino masses.
If neutrinos have normal mass hierarchy then a suppression factor 
arises so that one must go down to 0.001 eV level to probe the 
neutrino mass scale of $\sqrt{\Delta m^2_{atm}}$.

In my notation I assume that the 3rd mass eigenstate is either 
heaviest or lightest. I call the former (latter) case as the 
normal (inverted) mass hierarchy. I use as a lepton mixing matrix, 
the Maki-Nakagawa-Sakata matrix \cite {MNS}, 
the standard parametrization of the CKM matrix for quarks 
advertized by Partcle Data Group.
Then, the observable in double beta decay experiment can be written as
\begin{equation}
\langle m_{\nu e}\rangle = 
\left\vert c_{12}^2c_{13}^2 m_1 
+ s_{12}^2c_{13}^2 m_2 e^{i \phi_1}
+ s_{13}^2 m_3 e^{i \phi_2}
\right\vert,
\label{beta}
\end{equation}
where $\phi_1$ and $\phi_2$ are undetermined phases, essentially 
the two extra Majorana phases.
Since $\Delta m^2_{atm} = \Delta m^2_{23} \simeq \Delta m^2_{13}
\gg \Delta m^2_{12} = \Delta m^2_{solar}$,
$\langle m_{\nu e}\rangle$ is dominated by the 3rd (1st and 2nd)
term in (\ref{beta}) for the normal (inverted) mass hierarchy if 
I rely on the view represented in (\ref{numass}) with $M = M_{GUT}$.
An opposite extreme is known as the almost degenerate neutrinos
(ADN), $m_j^2 \gg \Delta m^2_{ij}$, and I refer an early analysis 
\cite {MY97} of the ADN scenario on how it can be constrained by 
the solar neutrino observation.

Now it is simple to observe that in the normal mass hierarchy
$\langle m_{\nu e}\rangle$ is given by 
\begin{equation}
\langle m_{\nu e}\rangle = 
s^2_{13} m_3 \simeq s^2_{13} \sqrt{\Delta m^2_{atm}}. 
\label{beta1}
\end{equation}
The constraint by the CHOOZ experiment \cite {CHOOZ}, 
$\sin^2 {2 \theta_{13}} \lsim 0.1$ and 
$\Delta m^2_{atm} = (2-5) \times 10^{-3} \mathrm{eV}^2$ gives 
$\langle m_{\nu e}\rangle \lsim (1.1-1.8)  \times 10^{-3}$ eV.
On the other hand, if the inverted mass hierarchy is the case 
\begin{equation}
\langle m_{\nu e}\rangle \simeq 
\left\vert c_{12}^2 m_1 
+ s_{12}^2 m_2 e^{i \phi_1}
\right\vert,
\label{beta2}
\end{equation}
which, barring the possibility of accidental cancellation, 
can be of the order of $\sqrt{\Delta m^2_{atm}}$, as announced 
in the abstract.

I don't know if I survived the Sphinx's questions. But what I told 
you in my talk is in what I believe. So let us wait and see.

In completing this manuscript I noticed that the similar 
issues on double beta decay have been addressed in \cite {KPS00}.

I thank Yoichiro Suzuki for asking such difficult questions and 
keep pressing us to answer.
This work is supported partly by the Grant-in-Aid for Scientific 
Research in Priority Areas No. 12047222 
Ministry of Education, Science, Sports and Culture.


\end{document}